\documentclass[prl,aps,showpacs,twocolumn,floatfix]{revtex4}
\usepackage{epsfig} \usepackage{graphics} \usepackage{bm}
\usepackage{amssymb}
\begin{document}

\preprint{Lebed-PRL}

\title{Hidden Reentrant and Larkin-Ovchinnikov-Fulde-Ferrell  Superconducting
Phases in a Magnetic Field in (TMTSF)$_2$ClO$_4$}

\author{A.G. Lebed$^*$}

\affiliation{Department of Physics, University of Arizona, 1118 E.
4-th Street, Tucson, AZ 85721, USA}

\begin{abstract}
We solve a long-standing problem about a theoretical description of the upper critical  magnetic field, parallel to conducting layers and perpendicular to conducting chains, in (TMTSF)$_2$ClO$_4$ superconductor.
In particular, we explain why the experimental upper critical field, $H^{b'}_{c2} \simeq 6 \ T$, is higher than both the quasi-classical upper critical field
and Clogston paramagnetic limit.
We show that this property is due to the coexistence of the hidden Reentrant and
 Larkin-Ovchinnikov-Fulde-Ferrell phases in a magnetic field in a form of
three plane waves with non-zero momenta of the Cooper
pairs.
Our results are in good qualitative and quantitative agreement  with the
recent experimental measurements of $H^{b'}_{c2}$ and support a singlet d-wave
like scenario of superconductivity in (TMTSF)$_2$ClO$_4$.

\end{abstract}


\maketitle


Physical properties of quasi-one-dimensional (Q1D) organic conductors
(TMTSF)$_2$X (X=PF$_6$, ClO$_4$, ReO$_4$, etc.) have been intensively
studied [1,2] since a discovery of superconductivity in
(TMTSF)$_2$PF$_6$ [3].
Early experiments [4,5] clear showed that superconducting phases in these
compounds were  unconventional and that the corresponding order parameters
changed their signs on Q1D Fermi surfaces (FS).
In particular, it was shown that the Hebel-Slichter peak was absent in the NMR experiment [4] and superconductivity was destroyed by non-magnetic
impurities [5].
These results have been recently confirmed in a number of publications (see, for example, Refs. [6,7]).
The first Knight shift measurements [7,8], performed in (TMTSF)$_2$PF$_6$
conductor in a magnetic field $H = 1.43 \ T$, showed that the Knight shift was
unchanged in superconducting phase and were interpreted as evidence
for triplet superconductivity.
On the other hand, more recent Knight shift data [9], performed in (TMTSF)$_2$ClO$_4$ conductor, clear demonstrate the Knight shift change
through the superconducting transition in a magnetic field
$H = 0.957 \ T$.
They are interpreted [9] in terms of singlet pairing in superconductor (TMTSF)$_2$ClO$_4$ at least at relatively weak magnetic fields.

Another source of information about a spin part of the superconducting order
parameter was provided by fact that the experimental upper critical magnetic
field along conducting chains, $H^a_{c2}$ [10], was clear paramagnetically
limited [11].
This has been recently confirmed in Refs. [12-14].
In addition, new superconducting phase has been discovered in (TMTSF)$_2$ClO$_4$ [12,13] for a magnetic field, parallel to conducting
chains.
The suggested hypothesis [12,13] that it can be the Larkin-Ovchinnikov-Fulde-Ferrell
(LOFF) phase [15,16] has been recently theoretically
supported [17].
Note that the above mentioned experimental and theoretical works
are in favor of a singlet d-wave like scenario of superconductivity in
(TMTST)$_2$ClO$_4$ [14,17-19] .

In this situation, where support for a singlet d-wave like scenario of superconducting pairing in (TMTST)$_2$ClO$_4$ conductor is growing, it is important theoretically
reinvestigate the upper critical field, parallel to conducting layers and
perpendicular to conducting chains, $H^{b'}_{c2}$.
For many years, large experimental values of $H^{b'}_{c2}$ [10,12,13,20-23],
which exceeds both the quasi-classical upper critical field, $H^{b'}_{c2}(0)$ [24],
and Clogston paramagnetic limit, $H_p$ [25], have been considered as one of the
main arguments in favor of
triplet superconductivity.
Although the exceeding of the values of  $H^{b'}_{c2}(0)$ and $H_p$
was predicted for $H^{b'}_{c2}$ in both singlet and triplet cases [26-30],
it was also shown that, for realistic band parameters of  (TMTST)$_2$X
conductors, it can happen only in a triplet case
[11,20,26-30].

The goal of our Letter is to demonstrate that superconductivity in (TMTSF)$_2$ClO$_4$
can exceed both critical magnetic fields, $H^{b'}_{c2} (0)\simeq 3.5 \ T$ and
$H_p \simeq 2.7 \ T$, and reach its experimental value, $H^{b'}_{c2} \simeq 6 \ T$ [12,13,23],
 even in case of a singlet d-wave like superconducting
pairing.
The first our point is that the Pauli paramagnetic effects in all previous theories
[26-30,11,20] were treated not completely correctly.
The second our point is that the $3D \rightarrow 2D$ dimensional crossover [26]
happens at magnetic fields $H^{b'} \simeq 5-6 \ T$, which are much lower than the previously assumed.
The latter statement is shown to result from theoretical analysis of both the Ginzburg-Landau (GL) slopes, $dH^{b'}_{c2}/dT|_{T=T_c}$ and $dH^{c}_{c2}/dT|_{T=T_c}$, measured in Refs.[12,13,23], and the so-called Lee-Naughton-Lebed oscillations [31,32].
In the Letter, we derive a novel gap equation, which treats accurately both the Pauli
paramagnetic and orbital destructive effects against superconductivity.
By analyzing this equation, we show that it predicts the upper critical field, $H^{b'}_{c2} \simeq 6 \ T$, for real values of band parameters in
(TMTSF)$_2$ClO$_4$.
Superconducting phase, which exists at such high magnetic fields, is shown
to be very peculiar.
It is characterized by a inhomogeneous order parameter in a form of the three
LOFF-like waves, which appear both due to the $3D \rightarrow 2D$ dimensional crossover and Pauli paramagnetic effects.
It is important that this phase is characterized by the Cooper pairs, localized on conducting layers, with probability of the Cooper pair jumping from one layer to
another being small.
Therefore, it is not destroyed by the orbital effects in a parallel magnetic
field.
In the absence of the Pauli paramagnetic effects, such phase would correspond
to the Reentrant superconductivity with $dT_c/dH >0$, therefore, we call it the
hidden Reentrant superconducting phase.

Below, we consider a tight-binding orthorhombic model of anisotropic
Q1D electron spectrum in (TMTSF)$_2$ClO$_4$ conductor,
\begin{equation}
\epsilon({\bf p})= - 2t_a \cos(p_x a/2) - 2 t_b \cos(p_y b') - 2t_c
\cos (p_z c^*),
\end{equation}
which can be simplified near two slightly corrugated sheets of Q1D FS as
\begin{equation}
\delta \epsilon^{\pm}({\bf p})= \pm v_x(p_y)[p_x \mp p_F(p_y)] - 2t_c \cos(p_z c^*) .
\end{equation}
[Here $t_a \gg t_b \gg t_c$ correspond to electron hoping integrals along
${\bf a}$ , ${\bf b'}$, and ${\bf c^*}$ axes, respectively; +(-) stands for right (left)
sheet of FS.]

In a magnetic field, parallel to conducting planes and perpendicular to conducting
chains of Q1D conductor,
\begin{equation}
{\bf H} = (0,H,0), \ \ {\bf A} = (0,0,-Hx),
\end{equation}
we use the so-called Peierls substitution method,
$p_x \mp p_F(p_y) \rightarrow -i d/dx$, $p_z \rightarrow p_z -eA_z/c$.
As a result, effective Scrodinger equation for electron wave functions in
a mixed representation, $\psi^{\pm}(x,p_y,p_z,\sigma)$,
can be written as
\begin{eqnarray}
\biggl[ \mp i v_x(p_y) \frac{d}{dx} &&- 2t_c \cos \biggl(p_zc^*+\frac{\omega_c}{v_F}x \biggl)
- \mu_B \sigma H \biggl] \nonumber\\
&&\times \psi_{\epsilon}^{\pm}(x,p_y,p_z,\sigma) =
 \delta \epsilon \ \psi_{\epsilon}^{\pm}(p_x,p_y,p_z,\sigma),
\end{eqnarray}
with electron wave in a real space functions being
\begin{eqnarray}
\Psi_{\epsilon}^{\pm}(x,y,z,\sigma) = \exp[i p_F(p_y)x] &&\exp(ip_y y) \exp(ip_z z)
\nonumber\\
&&\psi_{\epsilon}^{\pm} (x,p_y,p_z,\sigma),
\end{eqnarray}
where $\omega_c = ev_FHc^*/c$, $\mu_B$ is the Bohr magneton,
$\sigma=\pm1$ stands for spin up and down, respectively.

It is important that Eq.(4) can be analytically solved:
\begin{eqnarray}
&&\psi_{\epsilon}^{\pm}(x,p_y,p_z,\sigma)=
\frac{\exp[\pm i \delta \epsilon x /v_x(p_y)]}{\sqrt{2 \pi v_x(p_y)}}
\exp \biggl[\pm i \frac{\mu_B \sigma H x}{v_x(p_y)} \biggl]
\nonumber\\
&&\times \exp \biggl[ \pm i \frac{2t_c}{v_x(p_y)} \int_0^{x} \cos \biggl( p_z c^*
+ \frac{\omega_c}{v_F}u \biggl) du \biggl] ,
\end{eqnarray}
where wave functions (6) are normalized on $\delta(\epsilon_1-\epsilon_2)$,
$\delta \epsilon = \epsilon - \epsilon_F$.
The corresponding finite temperatures Green functions can be derived from Eq.(6)
by means of the standard procedure [33]:
\begin{eqnarray}
&&g^{\pm}_{i \omega_n} (x,x_1,p_y,p_z, \sigma)=
-i \frac{ sgn (\omega_n)}{v_x(p_y)}
\exp \biggl[ \mp \frac{\omega_n (x-x_1)}{v_x(p_y)} \biggl]
\nonumber\\
&&\times
\exp \biggl[ \pm i \frac{\mu_B \sigma H (x-x_1)}{v_x(p_y)} \biggl]
\nonumber\\
&&\times \exp \biggl[ \pm i \frac{2t_c}{v_x(p_y)} \int_{x_1}^{x} \cos \biggl( p_z c^*
+ \frac{\omega_c}{v_F}u \biggl) du \biggl] .
\end{eqnarray}
[Note that, in contrast to the previous works [26-30,11,20], Eqs.(6),(7) take
into account dependence of electron velocity along conducting chains, $v_x(p_y)$,
on a momentum component $p_y$.
As shown below, it allows to describe accurately the Pauli paramagnetic
destructive effects against superconductivity.]

In this Letter, we consider a singlet d-wave like scenario of superconductivity in
(TMTSF)$_2$ClO$_4$ conductor [14,17-19], which is consistent with all available
experimental data.
Therefore, we introduce the following superconducting order parameter,
\begin{equation}
\Delta (p_y,x) = \sqrt{2} \cos(p_y b') \Delta (x),
\end{equation}
where the first term, $\sqrt{2} \cos(p_y b')$, is responsible for the existence of zeros
on Q1D FS, whereas the second term describes both the orbital effects against superconductivity and possible LOFF like phase
formation.
Below, we derive a so-called gap equation for the superconducting order
parameter (8), using the Green functions (7).
It is derived by means of the Gor'kov equations [33] for non-uniform superconductivity
(see, for example, Refs. [34-36].
As a result of rather lengthly calculations, we obtain:
\begin{eqnarray}
&&\Delta(x) = g'  \int \frac{d p_y}{v_x(p_y)} \int_{|x-x_1| > \frac{v_x(p_y)}{\Omega}}
\frac{2 \pi T dx_1}{v_x(p_y) \sinh \bigl( \frac{2 \pi T |x-x_1|}{v_x(p_y)} \bigl)}
\nonumber\\
&&\times J_0 \biggl\{ \frac{8 t_c v_F}{\omega_c v_x(p_y)}
\sin \biggl[ \frac{\omega_c (x-x_1)}{2v_F} \bigg]
\sin \biggl[ \frac{\omega_c (x+x_1)}{2v_F} \bigg]
\biggl\}
\nonumber\\
&&\times 2 \cos^2 (p_y b') \cos \biggl[ \frac{2 \beta \mu_B H (x-x_1)}{v_x(p_y)} \biggl] \
\Delta (x_1) ,
\end{eqnarray}
where $g'$ stands for electron coupling constant, $\Omega$ is a
cutoff energy, parameter $\beta$ takes into account possible
deviation of the so-called electron $g$-factor, $g = 2 \beta$, from
the value $g=2$ [37]. We stress that Eq.(9) is different from the
gap equations, used so far, and, unlike Refs. [26-30,11,20], it
describes accurately not only the orbital effects but also the Pauli
paramagnetic ones. Note that Eq.(9) is based on a quantum mechanical
treatment of electron motion both in parallel and perpendicular to
conducting layers directions. It is the most general gap equation,
which can be written for Q1D conductor (2) in a magnetic field (3).
It is possible to show that the major quantum parameter in Eq.(9) is
$2t_cv_F/\omega_cv_x(p_y) \simeq 2t_c/\omega_c$. It is also possible
to prove that in low magnetic fields, where $2t_c/\omega_c \gg 1$
and $(T_c-T)/T_c \ll 1$, Eq.(9) is reduced to the well known GL
equation [38].

Let us estimate a value of the dimensionless quantum parameter
$l_{\perp}(H)=2t_c/\omega_c$ in Eq.(9), which, using classical
language, represents a size of electron trajectory along ${\bf z}$
axis in terms of interlayer distance [26]:
\begin{equation}
z(t,H) = l_{\perp}(H) \ c^* \cos(\omega_c t) ,
\end{equation}
where $t$ is time.
It is easy to show that
\begin{equation}
l_{\perp}(H) = \frac{2 \sqrt{2}}{\pi} \frac{\phi_0}{a c^* H} \frac{t_c}{t_a} \simeq
\frac{2 \times 10^3}{H(T)} \frac{t_c}{t_b} \frac{t_b}{t_a} ,
\end{equation}
where $H(T)$ is a magnetic field, measured in Teslas. Here,
according to Ref.[32], $t_a/t_b = 10$ and, according to Ref.[38],
$t_c/t_b= (b^*/\sqrt{2} c^*) (H^c_{c2}/H^{b'}_{c2})_{GL}$ with
$(H^c_{c2}/H^{b'}_{c2})_{GL}$ is being a ratio of the GL slopes of
the upper critical fields along $c^*$ and $b'$ axes, correspondingly
[39].
Note that the ratios $t_a/t_b =10$ [31] and
$(H^{b'}_{c2}/H^{c}_{c2})_{GL}=26$ [12,13] are very well measured in
(TMTSF)$_2$ClO$_4$ conductor. If we take $H(T)=6 \ T$, we obtain
\begin{equation}
l_{\perp}(H=6T) \simeq 0.48 ,
\end{equation}
which means that a size of electron classical trajectory along ${\bf
c^*}$ axis (10) is significantly less than interlayer distance,
$c^*$. In this case, which corresponds to the $3D \rightarrow 2D$
dimensional crossover of electron motion in a magnetic field
[26,40], it is possible to make sure directly from Eq.(9) that we
can approximate the Bessel function as $J_0(z) \simeq 1 - z^2/4$.

Let us consider the above mentioned approximation for integral equation (9)
at zero temperature, $T=0$.
It is possible to show that solution for a superconducting gap, $\Delta(x)$, in
this case can be written as
\begin{equation}
\Delta (x) = \exp(ikx)[1 + \alpha_1 \cos(2 \omega_c x /v_F)
+ \alpha_2 \sin(2 \omega_c x /v_F) ],
\end{equation}
where $|\alpha_1|, |\alpha_2| \ll 1$.
Eq.(9), determining the upper critical field, in the same approximation
and at $T=0$ can be expressed as
\begin{eqnarray}
\frac{1}{\tilde{g}} = \int^{2 \pi /b'}_{0} \frac{d p_y b'}{2 \pi }  \int^{\infty}_{\frac{v_F}\Omega} \frac{d z}{z}
2 \cos^2(p_y b') \cos \biggl( \frac{2 \beta \mu_b H z}{v_F} \biggl)
\nonumber\\
\times  \frac{v_F}{v_x(p_y)}  \biggl[1-2 l^2_{\perp}(H) \sin^2 \biggl(\frac{\omega_c z}{2 v_F} \biggl) \biggl] \cos \biggl[  \frac{v_x(p_y)}{v_F} k z \biggl]  ,
\end{eqnarray}
where $\tilde{g}$ is renormalized electron coupling constant,
$x_1-x=z v_x(p_y)/v_F$.
[Note that we set $\alpha_1= \alpha_2 =0$ in Eq.(14), since we disregard
all contributions of the order of $l^4_{\perp}(H) \ll l^2_{\perp}(H)$ to the
upper critical field.]

Below, we simplify Eq.(14), taking into account that electron velocity
component along conducting ${\bf x}$ axis is
\begin{equation}
v_x(p_y) = v_F [1 + \alpha \cos(p_y b')] ,
\end{equation}
where $\alpha = \sqrt{2} t_b / t_a \simeq 0.14$ [20].
More specifically, Eq.(14) for $\alpha \ll 1$ can be written as follows:
\begin{eqnarray}
\frac{1}{\tilde{g}}= \int^{\infty}_{\frac{v_F}{\Omega}} \frac{dz}{z}
&&\cos \biggl( \frac{2 \beta \mu_B H z}{v_F} \biggl) \cos(k z) [J_0(\alpha k z) - J_2(\alpha kz)]
\nonumber\\
&&\times \biggl[1-2 l^2_{\perp}(H) \sin^2 \biggl(\frac{\omega_c z}{2 v_F} \biggl) \biggl] .
\end{eqnarray}
It is important that Eq.(16) accurately takes into account the Pauli paramagnetic
effects against superconductivity,
unlike Refs.[26-30,11,20].
Note that, in the absence of the Pauli paramagnetic effects (i.e., at $\beta =0$),
Eq.(16) describes the Reentrant superconducting phase [26] with
$dT_c/dH>0$.
Therefore, we call superconducting phase, described by Eqs.(16),(18),
the hidden Reentrant superconductivity.

Let us further simplify Eq.(16) by taking into account that
\begin{equation}
\frac{1}{\tilde{g}} = \int^{\infty}_{\frac{v_F}{\Omega}}
\frac{2 \pi T_c dz}{v_F \sinh \biggl( \frac{2 \pi T_c z}{v_F} \biggl)} ,
\end{equation}
where $T_c$ is the superconducting transition temperature at $H=0$.
As a result, we obtain
\begin{eqnarray}
&&\ln \biggl (\frac{H}{H^*} \biggl) = \int^{\infty}_{0} \frac{dz}{z} \cos \biggl( \frac{2 \beta \mu_B Hz}{v_F}
\biggl)
\nonumber\\
&&\times \biggl\{\cos(kz) [J_0(\alpha k z)-J_2(\alpha k z)]
\nonumber\\
&&\times \biggl[1 - 2 l^2_{\perp}(H) \sin^2 \biggl( \frac{\omega_c z}{2 v_F} \biggl) \biggl] -1 \biggl\} ,
\end{eqnarray}
where $\mu_B H^* = \pi T_c / 2 \gamma$,
$\gamma$ is the Euler constant.
Numerical analysis of Eq.(18) shows that the upper critical field along ${\bf b'}$ axis, $H^{b'}_{c2}$, for $l_{\perp}(H)=0.48$ and $\beta=0.84$ has a maximum at
$k=0.88 (2 \beta \mu_B H / v_F)$ and  is equal to
\begin{equation}
H^{b'}_{c2} \simeq 5.9 \ T .
\end{equation}
[We pay attention that the obtained value of the upper critical field (19) well
corresponds to the value of a magnetic field (12).]
For the same values of the parameters $l_{\perp}(H)$ and $\beta$, numerical
analysis of Eq.(9) gives the following values for factors $\alpha_1$ and $\alpha_2$
in Eq.(13):
\begin{equation}
\alpha_1= -0.139 , \ \ \ \alpha_2= 0.021 \ i .
\end{equation}

Below, we summarize the main results of the Letter. We have derived
gap equations (9),(14),(16),(18), which, unlike gap equations in the
previous publications, take accurately into account  not only the
orbital effects, but also the Pauli paramagnetic effects against
superconductivity. We have analyzed the experimental data [12,13,32]
and shown that, in contrast to the common believe, the quantum
effects of electron motion in a magnetic field [26,41] are strong in
relatively weak magnetic fields of the order of $5-6 \ T$ in
(TMTSF)$_2$ClO$_4$ conductor. By analyzing the above mentioned gap
equations, we have explained how superconductivity in
(TMTSF)$_2$ClO$_4$ can exceed both the quasi-classical upper
critical field [24] and Clogston paramagnetic limit [25] and how it
can reach its experimental value, $H \simeq 6 \ T$ [12,13]. We have
shown that, due to the reentrant quantum effects [26,41],
superconductivity  survives in form of the hidden Reentrant
superconducting phase, corresponding to three LOFF-like phases.
Although we have not calculated in the Letter phase diagram of the
(TMTSF)$_2$ClO$_4$ superconductor in all range of temperatures and
magnetic fields, we anticipate the existence of phase transition
between the BCS and LOFF phases at $H \simeq 2.5 \ T$, which can be
experimentally studied.

In conclusion, we note that the considered above hidden Reentrant
superconductivity is a rather general phenomenon. It is expected to
exist in other (TMTSF)$_2$X conductors and may exist in
quasi-two-dimensional superconductors in a parallel magnetic field.
Nevertheless, this phase in (TMTSF)$_2$PF$_6$ material, which is
stable in a mixed superconducting-spin-density-wave state [21,22,42]
in a magnetic field up to $H= 9 \ T$, possesses some peculiarities.
Our preliminary analysis shows that, to describe the hidden
Reentrant superconducting phase in (TMTSF)$_2$PF$_6$, it is
necessary to take into account some additional effects such as the
singlet-triplet mixing phenomenon [43] or possible singlet-triplet
phase transition (see, for example, [30,20]).

We are thankful to N.N. Bagmet for useful discussions. This work was
supported by the NSF under Grant No DMR-0705986.

$^*$Also at the Landau Institute for Theoretical Physics,
2 Kosygina Street, Moscow 117334, Russia.


\begin{references}
\bibitem{Lebed-51}
It also accounts for possible small deviations of superconducting pairing
in (TMTSF)$_2$ClO$_4$ conductor from a weak coupling scenario.
\bibitem{Lebed-61}
Note that, in our calculations, we use parameter $t_b/t_c$,
determined from the GL slopes of the anisotropic upper critical
field, measured near $T_c \simeq 1.45 \ K$. We disregard possible
small changes of this parameter at lower temperatures.
\bibitem{Lebed-62}
We have proven that, in the absence of the Pauli paramagnetic
effects, Eq.(9) at $l_{\perp}(H)=0.48$ leads to the Reentrant
superconductivity with $dT_c/dH>0$, which we interpret as $3D
\rightarrow 2D$ dimensional crossover.
\end{references}
\end{document}